\documentclass{PoS}

\usepackage{bm}
\usepackage{lineno}

\title{Higgs boson measurements and extended scalar sector searches in fermionic final states}

\ShortTitle{Higgs boson measurements and searches in fermionic final states}

\author{\speaker{Teresa Lenz} (on behalf of the CMS Collaboration)\\
        DESY, Notkestr. 85, 22607 Hamburg, Germany\\
        E-mail: \email{teresa.lenz@desy.de}}

\abstract{Since the discovery of the Higgs boson in the year 2012, the CMS experiment continuously makes use of a variety of final states to measure the properties of the Higgs boson, 
          as well as to search for it in final states where the discovery could not yet be established.
          This article concentrates on Higgs searches in fermionic final states including a search for the Standard Model Higgs boson in the $H \rightarrow b\overline{b}$ channel.
          Furthermore, three different searches for Higgs bosons beyond the Standard Model are presented.
          These comprise two searches for heavy neutral Higgs bosons and one search for a heavy charged Higgs boson, all of them exploiting final states with tau leptons.
          }

\FullConference{XXV International Workshop on Deep-Inelastic Scattering and Related Subjects\\
		3-7 April 2017\\
		University of Birmingham, UK}

\begin{document}


\section{Search for the Standard Model Higgs boson}
With data recorded in the year 2012 at a center-of-mass energy of 8\,TeV, the CMS experiment~\cite{CMS_experiment} was able to discover the Higgs boson in many different final states. 
All of these measurements are in agreement with Standard Model (SM) predictions (see Fig.~\ref{fig:HiggsXS}\,(left)).
However, despite being the final state with the highest branching fraction, the decay channel $H \rightarrow b\overline{b}$ is still open for discovery.
This is caused by the overwhelming QCD-multijet background which makes this decay channel very hard to detect.
The most promising production channel to discover $H \rightarrow b\overline{b}$ is the vector boson fusion (VBF) channel.
Therefore, the search for the Higgs boson decaying into two bottom quarks and produced in association with two jets is presented in the following section.

\subsection{Search for the Standard Model Higgs boson produced through vector boson fusion and decaying to $\mathbf{b\overline{b}}$ at $\mathbf{\sqrt{s} = 13\,TeV}$}
\begin{figure}[b]
 \centering
 \begin{minipage}{0.36\textwidth}
 \centering
 \includegraphics[width=0.99\textwidth]{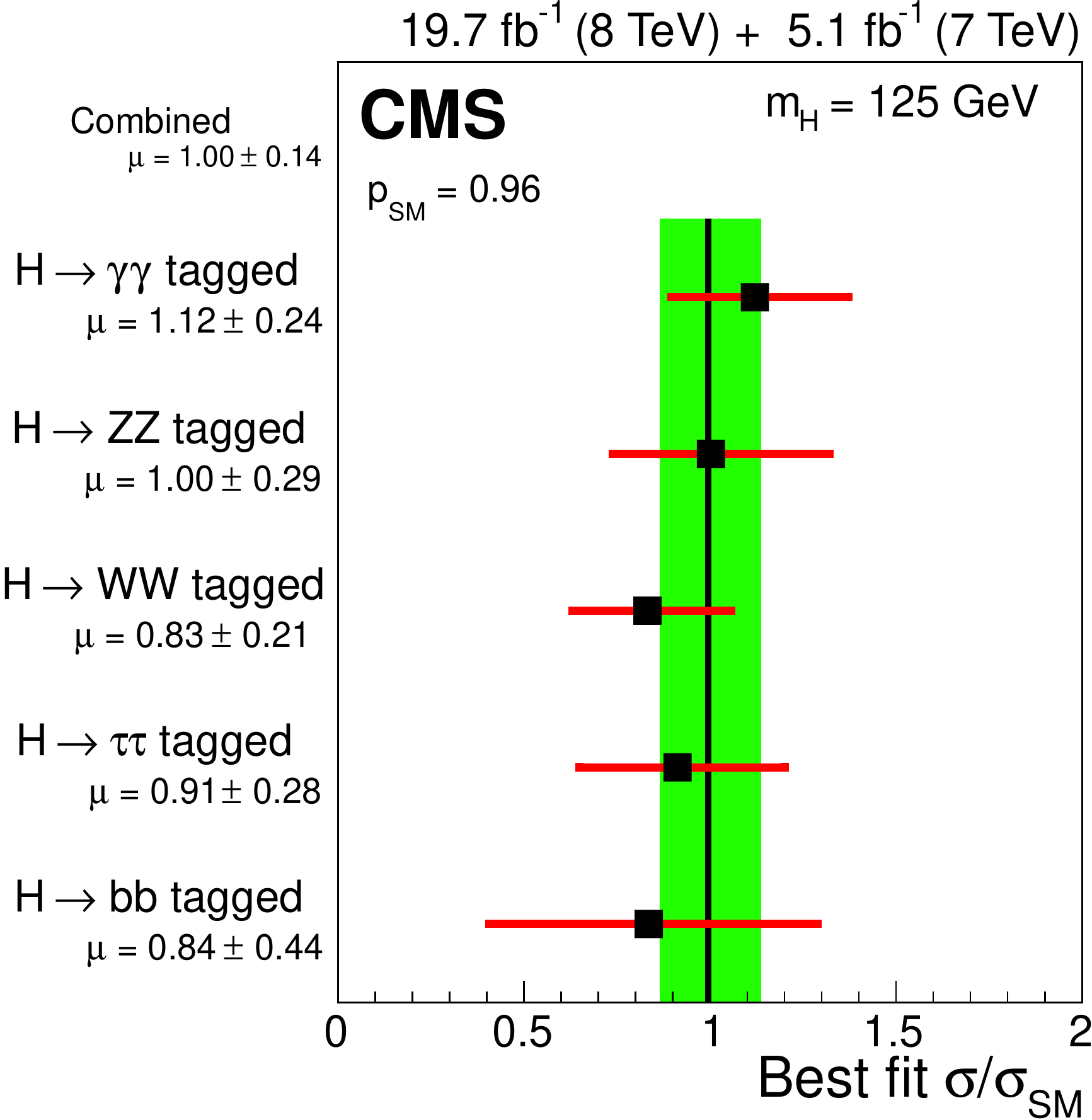}
 \end{minipage}
 \hspace{2cm}
 \begin{minipage}{0.36\textwidth}
 \centering
 \includegraphics[width=0.99\textwidth]{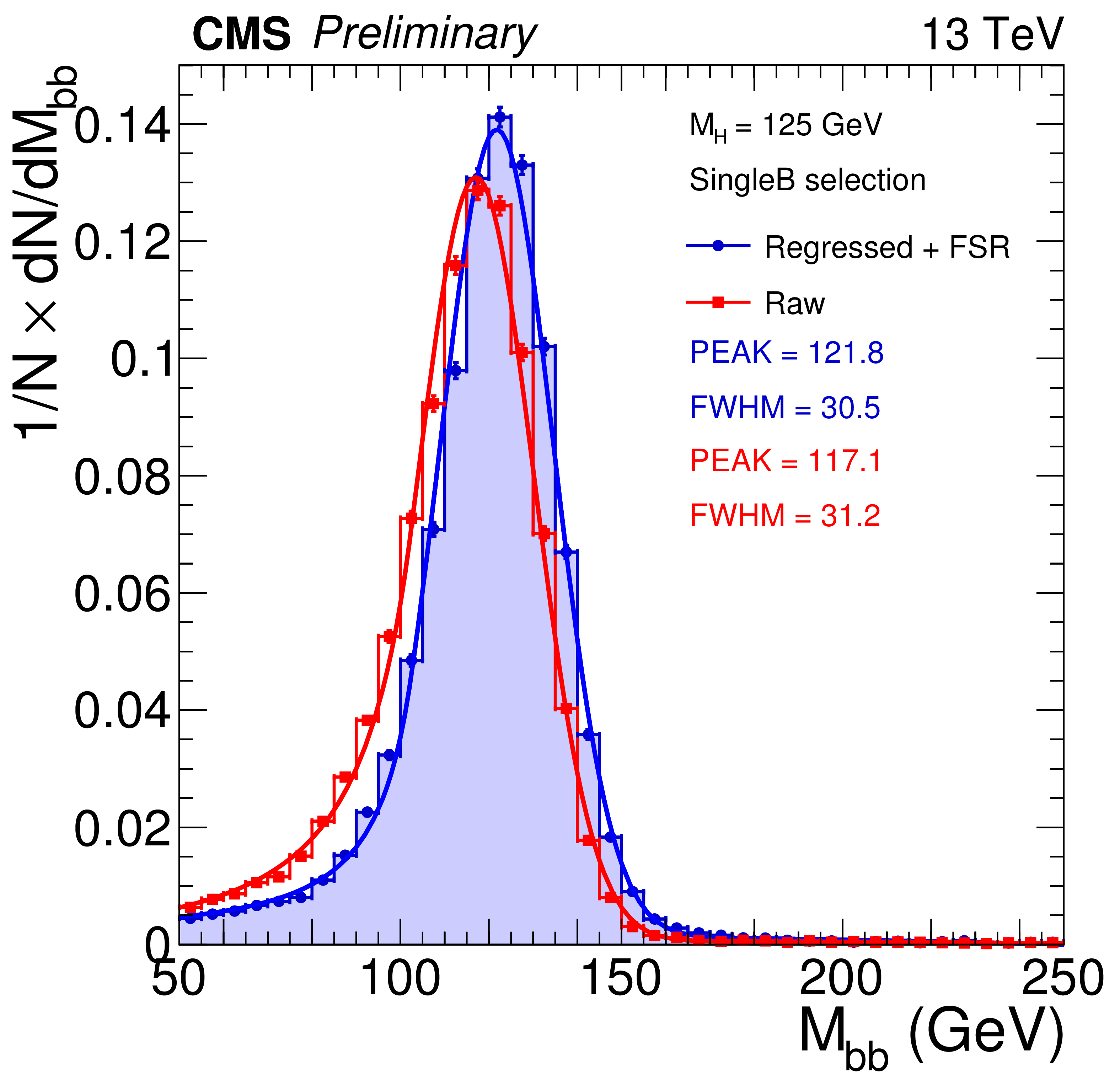}
 \end{minipage}
 \caption{Left: Best-fit $\sigma/\sigma_{\mathrm{SM}}$ for different predominant decay modes~\cite{CMS_2012_Summary}. 
          Right: Simulated invariant mass of the two b quark jets before (red) and after (blue) the improvements explained in the text~\cite{CMS_2015_Hbb}.}  
 \label{fig:HiggsXS}
\end{figure}

The characteristic feature of $H \rightarrow b\overline{b}$ decays in the vector boson fusion channel~\cite{CMS_2015_Hbb} are four energetic jets in the final state,
two of which are b quark jets and two are forward jets.
Since the final state consists exclusively of jets, the main challenge of this search is the separation of signal events against the QCD-multijet background.
Since the Higgs is produced via $W$ or $Z$ bosons, the production is a fully electroweak process and therefore no color connection between the final state jets exists.
This special feature of the Higgs production in this channel can be used to separate signal and QCD background events.
A further challenge is the reconstruction of the Higgs boson mass from the two b quark jets since semi-leptonic b decays come along with neutrinos 
and lead to loss of information of the full energy of the b~quarks.
Improvements with respect to the 8\,TeV analysis are achieved by determining b quark jet specific jet energy corrections which are derived with multivariate regression techniques. 
Additionally, methods to recover gluon radiation that is not clustered into the jet cone are exploited.
These two additions lead to an improvement of the jet mass resolution of up to 7\% (cf. Fig.~\ref{fig:HiggsXS}\,(right)).

The final statistical inference is done with the help of multivariate analysis techniques.
A boosted decision tree (BDT) is used to classify events into signal and background events.
Various observables are used for the classification, like kinematic variables of jets, b-jet discriminators, QCD gap activity, etc.
From this set of input variables, the BDT builds a test statistics called BDT output which is used to build four different categories which help to improve the sensitivity of the search.
As final observable, the invariant mass of the two b quark jets is used to estimate the best-fit signal strength.
This analysis makes use of 2.3\,fb$^{-1}$ of collected data at 13\,TeV.
The signal strength relative to the SM expectation is estimated to a negative value of $\mu = \sigma/\sigma_{\mathrm{SM}} = -3.7^{+2.4}_{-2.5}$ 
which is below zero because of a statistical downward fluctuation of the background contributions.
Combining this result with the result from the 8\,TeV analysis, a signal strength of $\mu = 1.3^{+1.2}_{-1.1}$ with an observed (expected) signal significance of $1.2\sigma$ ($0.9\sigma$) is found.

\section{Searches for Higgs bosons beyond the Standard Model}

In many theories beyond the Standard Model (BSM), multiple Higgs bosons are present. 
This is for example the case in supersymmetric extensions of the Standard Model where at least two Higgs doublets are needed in order to give masses to all particles.
In the Minimal Supersymmetric Standard Model (MSSM) five physical Higgs bosons exist: two CP-even ($h,\, H$), one CP-odd ($A$), and two charged Higgs bosons ($H^{\pm}$).
The lightest CP-even state $h$ is usually associated with the $h( 125 \mathrm{\,GeV})$ state.
In the MSSM, there are only two relevant parameters at tree level which fully describe the Higgs sector: the ratio of the vacuum expectation values of the two Higgs doublets, $\tan\beta$, 
and the mass of the CP-odd Higgs boson, $m_A$.
In the following sections, three different searches for neutral and charged heavy Higgs bosons are presented.

\subsection{Search for a neutral MSSM Higgs boson decaying into $\tau\tau$ with 12.9\,fb$^{-1}$ of data at $\mathbf{\sqrt{s} = 13\,TeV}$}

Since Higgs couplings are proportional to the mass of fermions and, furthermore, in the large $\tan\beta$ region the couplings to down-type fermions are enhanced, 
searching in final states with b~quarks or tau leptons is especially promising.
Furthermore, because of the better discrimination against SM processes like QCD-multijet events, 
the di-tau final state is the most sensitive final state to search for neutral heavy Higgs bosons in the large $\tan\beta$ region. 
A search for a neutral MSSM Higgs boson decaying into two tau leptons was performed at CMS with 12.9\,fb$^{-1}$ of data at 13\,TeV~\cite{CMS_2016_HTauTau_BSM}.

The event selection of this search includes the selection of two tau leptons where four out of six possible final states are used 
($\tau_h\tau_h$, $\tau_h\tau_e$, $\tau_h\tau_{\mu}$, $\tau_e\tau_{\mu}$).
Furthermore, two different categories are defined, one category with no b-tagged jets and one category with at least one b-tagged jet.
This helps in improving the search sensitivity especially in the large $\tan\beta$ region where the b associated production becomes more important.
The four different final states differ in their background composition and are therefore optimized and estimated separately and only combined in the statistical inference of the analysis. 

The final observable is the total transverse mass which is defined as the square root of the quadrature sum of all combinations of 
the transverse momentum of the final state leptons and the missing transverse energy of an event
$$ m_{\mathrm{T}}^{\mathrm{tot}} = \sqrt{m_{\mathrm{T}}\left( E_{\mathrm{T}}^{\mathrm{miss}},\tau_1^{\mathrm{vis}} \right)^2 + m_{\mathrm{T}}\left( E_{\mathrm{T}}^{\mathrm{miss}},\tau_2^{\mathrm{vis}} \right)^2 + m_{\mathrm{T}}\left(\tau_1^{\mathrm{vis}} ,\tau_2^{\mathrm{vis}} \right)^2 }.$$
Only the transverse momentum of the visible components of the tau lepton decay products are used for this variable.

No excess above the Standard Model prediction is found and therefore exclusion limits in the $m_A - \tan\beta$ plane are set (see Fig.~\ref{fig:Limits}\,(left)).
This search provides the first exclusions that extend beyond $m_A > 1\,$TeV at CMS.

\subsection{Search for charged Higgs bosons with the $\mathbf{H^{\pm} \rightarrow \tau^{\pm} \nu_{\tau}}$ decay channel in the fully hadronic final state at $\mathbf{\sqrt{s} = 13\,TeV}$}

Because of the additional Higgs doublet, two charged Higgs bosons are present in the MSSM.
A search for charged heavy Higgs bosons in final states with a tau lepton and a tau lepton neutrino was performed at CMS with 12.9\,fb$^{-1}$ of data at 13\,TeV~\cite{CMS_2016_HTauNu_BSM}.
The production mechanism depends on the mass of the charged Higgs boson.
If the Higgs mass is smaller than the top mass, the Higgs boson is produced in decays of the top quark, 
while in case the Higgs boson is heavier than the top quark it is produced in association with a top and a b quark.
Still, the final state consists of the same particles for both cases ($2b$, $2q$, $1\tau$, $1\nu_{\tau}$).
Therefore, this search selects events with one tau lepton in the final state ($p_{\mathrm{T}}^{\tau}>50$\,GeV), missing transverse energy ($>90$\,GeV), 
and at least three jets, one of which must be b-tagged.
It additionally requires that the tau lepton and the missing transverse energy do not show a back-to-back topology which leads to a suppression of QCD-multijet events.

As final observable the transverse mass between the tau lepton and the missing transverse energy is used which is expected to be close to the mass of the charged Higgs boson for a positive signal.
No deviation from SM predictions is found and therefore exclusion limits are set in the $m_{H^{\pm}}-\tan\beta$ plane for the $m_h^{\mathrm{mod+}}$ scenario (see Fig.~\ref{fig:Limits}\,(center)).
For the high mass scenario \mbox{($m_{H^{\pm}} > m_{t}$)} models with Higgs masses up to $m_{H^{\pm}} \approx 450$\,GeV are excluded.
Model independent upper limits on cross section times branching ratio,
$\sigma \left( pp \rightarrow H^{\pm}W^{\mp}b\overline{b} \right) \cdot B\left( H^{\pm} \rightarrow \tau^{\pm} \nu_{\tau}  \right)$,
reach up to~$\approx2$\,pb for $m_{H^{\pm}} =180$\,GeV.

\subsection{Search for pair production of Higgs bosons in the two tau leptons and two bottom quarks final state at $\mathbf{\sqrt{s} = 13\,TeV}$}

The resonant pair production of the Higgs boson at 125\,GeV is possible in many models beyond the SM, including the MSSM.
In the MSSM, the resonance can be caused by the heavy neutral CP-even Higgs boson $H$.
It is possible to search in a variety of final states for this process depending on the decay channel of the Higgs boson. 
This article presents a search for resonant Higgs pair production in final states with two tau leptons and two b quarks~\cite{CMS_2016_HToHHToTauTauBB_BSM}.

If the mass of the heavy Higgs boson is larger than $\gtrsim 700$\,GeV, the two b quark jets start to overlap.
Therefore, the two b jets are additionally reconstructed as one large-cone jet besides their reconstruction as two small-cone jets.
This special feature allows a clear separation against $t\overline{t}$~background where the two b jets are usually farther apart.
In order to enhance the sensitivity in this high mass region, a ``boosted category'' is introduced.

To separate signal against background events, mass reconstructions are performed for the invariant masses of the two b quark jets and the two tau leptons which are expected to be close to 125\,GeV for signal events.
Using these reconstructions, a third mass reconstruction is performed which aims at reconstructing the heavy Higgs mass $m_{H}$.
This variable is used for the final statistical analysis of this search.

\begin{figure}[t]
 \centering
 \begin{minipage}{0.32\textwidth}
 \centering
 \includegraphics[width=0.99\textwidth]{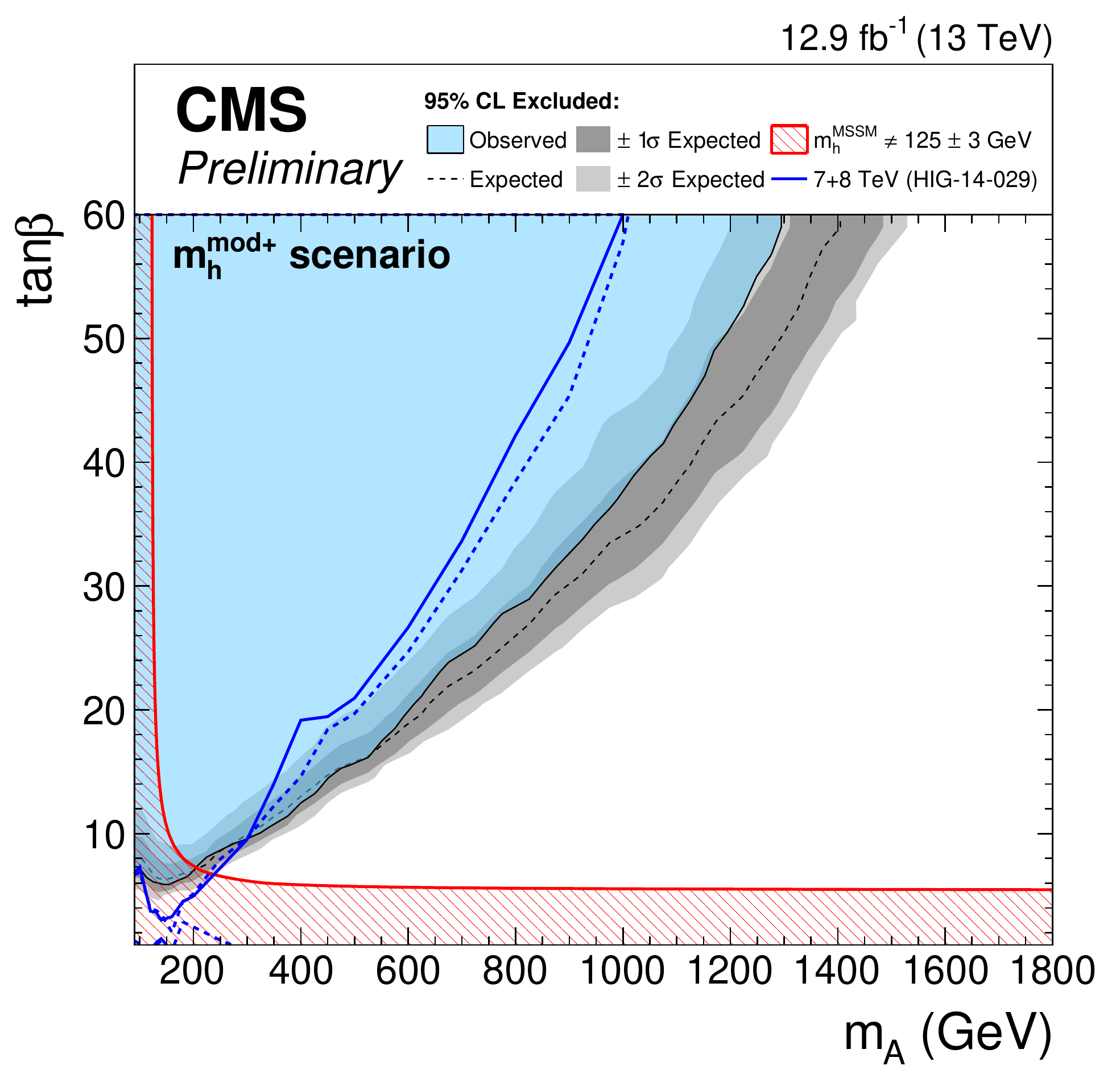}
 \end{minipage}
 \hfill
 \begin{minipage}{0.32\textwidth}
 \centering
 \includegraphics[width=0.99\textwidth]{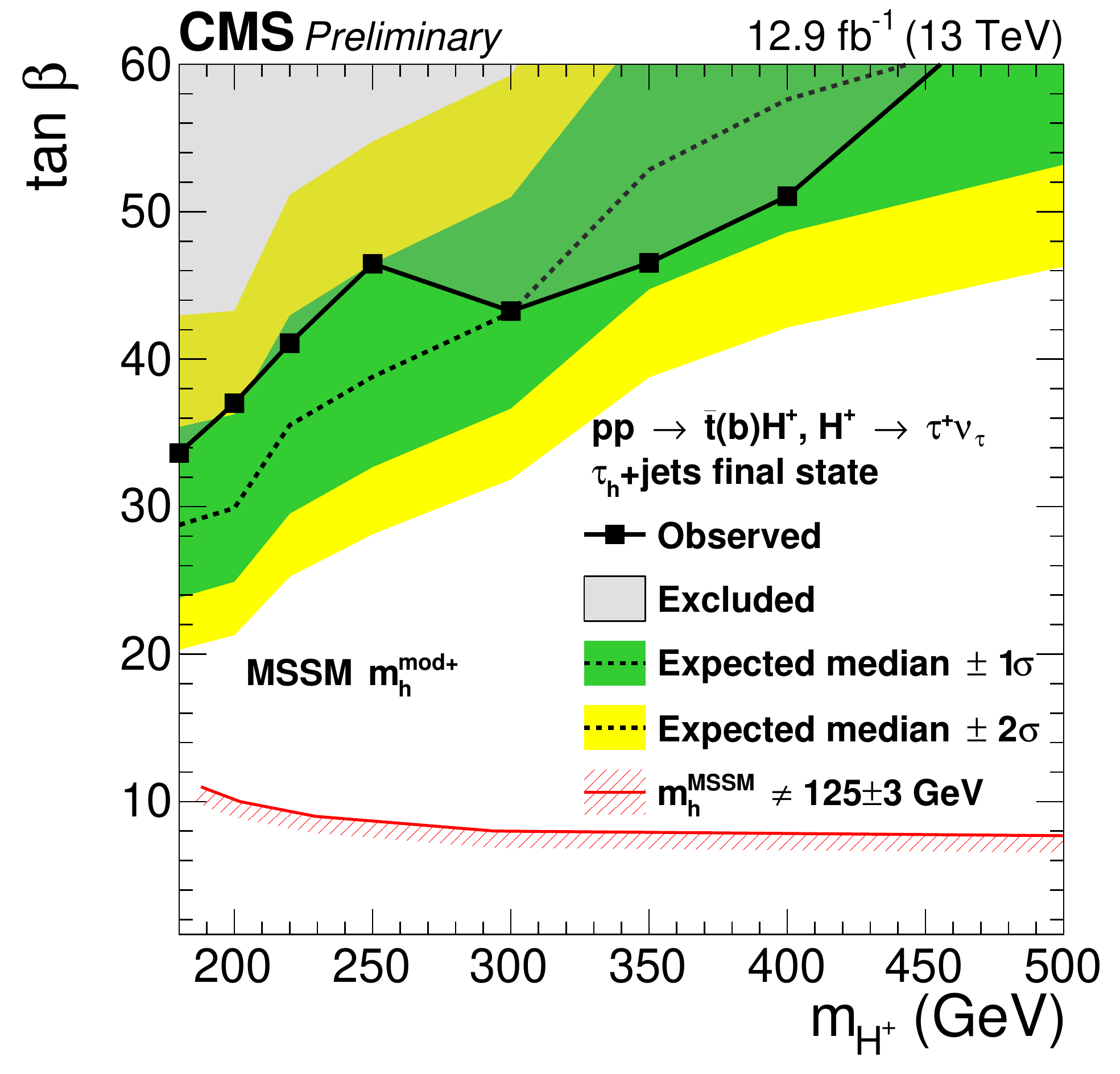}
 \end{minipage}
 \hfill
 \begin{minipage}{0.32\textwidth}
 \centering
 \includegraphics[width=0.99\textwidth]{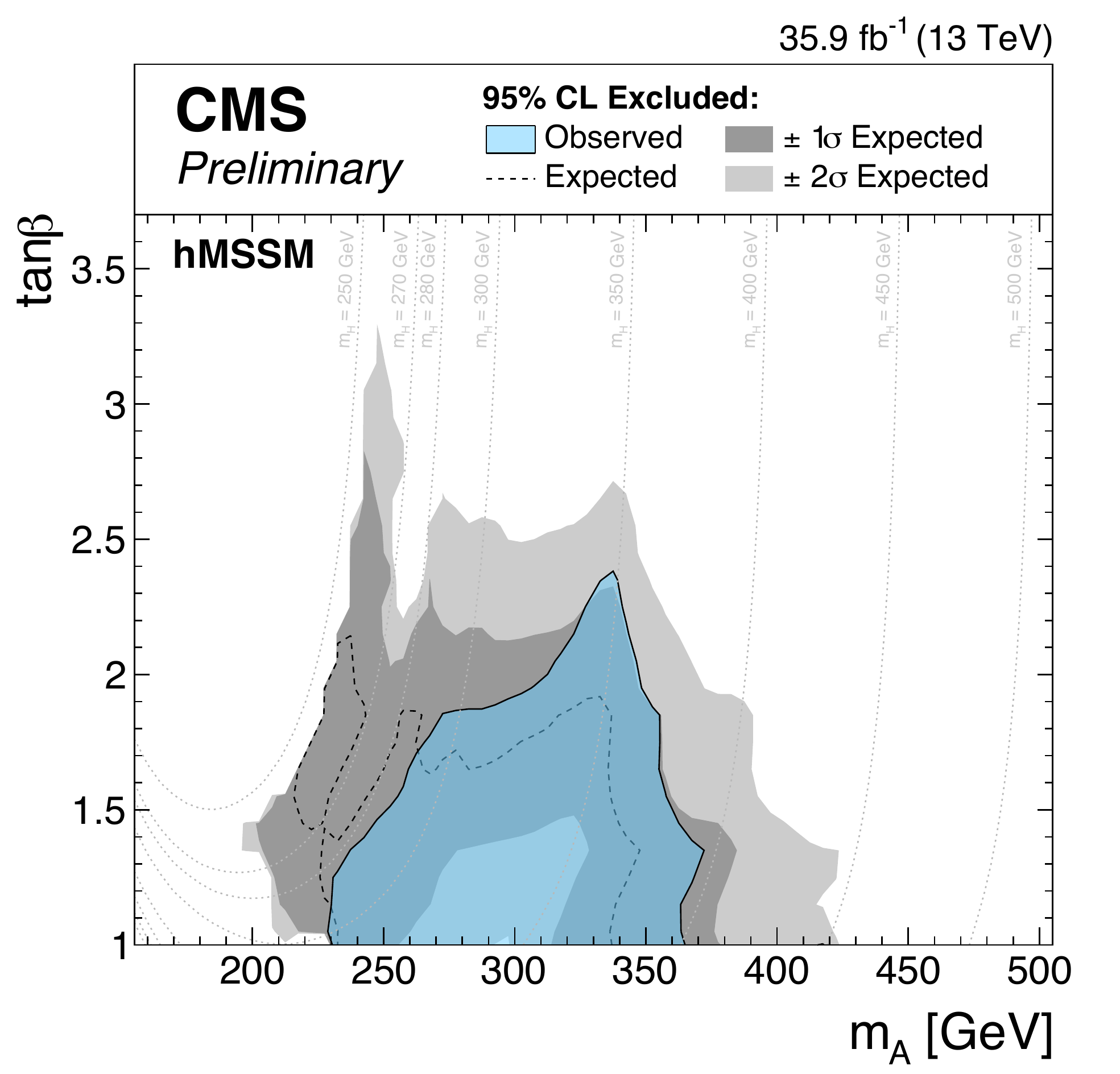}
 \end{minipage}
 \caption{Exclusion limits of the BSM $H\rightarrow\tau\tau$ search~\cite{CMS_2016_HTauTau_BSM} (left), the $H^{\pm}\rightarrow\tau\nu_{\tau}$ search~\cite{CMS_2016_HTauNu_BSM} (center),
          and the BSM $H \rightarrow h(125) h(125) \rightarrow b b \tau \tau $ search~\cite{CMS_2016_HToHHToTauTauBB_BSM} (right).}  
 \label{fig:Limits}
\end{figure}

Since no excess above background expectations is found, the result is interpreted in the context of the MSSM 
and exclusion curves in the $m_A$ versus $\tan\beta$ plane are derived (see Fig.~\ref{fig:Limits}\,(right)).
Model independent 95\% confidence level upper limits on the cross section times branching ratio, $\sigma\left( pp \rightarrow H \right)\cdot$ $B\left( H \rightarrow h h \rightarrow b b \tau \tau  \right)$,
reach up to $\approx600$\,fb for $m_{H} \approx 270$\,GeV.

\section{Conclusion}
At 8\,TeV, the SM Higgs boson was established in a variety of final states at the CMS experiment.
The only missing final state not yet showing evidence for the SM Higgs boson is the $H \rightarrow b\overline{b}$ decay mode.
The first search at 13\,TeV for $H \rightarrow b\overline{b}$ in the VBF channel is compatible with SM expectations, though does not improve the significance compared to the 8\,TeV analysis 
because of an under fluctuation of the background.
Combining the 8 and 13\,TeV analyses lead to a best-fit signal strength of $\mu = 1.3^{+1.2}_{-1.1}$ with an observed (expected) signal significance of~$1.2\sigma$ ($0.9\sigma$).

Furthermore, Higgs boson searches beyond the Standard Model do not show any evidence for an extended Higgs sector.
However, the MSSM parameter space is further constrained because of the significant sensitivity increases in the $m_A - \tan\beta$ plane with respect to the 8\,TeV analyses.

\section*{ACKNOWLEDGMENTS}
I thank the CMS collaboration for the work presented here, the CERN accelerator division for the excellent operation of the LHC, and all funding agencies that made this experiment possible.

\end{document}